\documentclass[seceq,preprint]{ptptex}

\usepackage{graphicx}
\usepackage{bm}

\newcommand{\rbk}[1]{\left( #1 \right)}

\newcommand{\abs}[1]{\left| #1 \right|}
\newcommand{\nbk}[1]{\left. #1 \right.}
\newcommand{\retn}{\nonumber \\ }

\newcommand{\dif}[2]{\frac{\mathrm{d} #1}{\mathrm{d} #2}}

\newcommand{\pdif}[2]{\frac{\partial #1}{\partial #2}}

\newcommand{\vc}[1]{\bm{#1}}
\newcommand{\ra}[0]{\rightarrow}

\notypesetlogo                       

\title{
Energy Current with Multi--body Interaction using Dirac Delta Function%
}

\author{
Atsushi \textsc{Ito}$^1$ and Hiroaki \textsc{Nakamura}$^2$%
}

\inst{
$^1$Department of Physics, Graduate School of Science, Nagoya University, Furo-cho, Chikusa-ku, Nagoya 464-8602, Japan.
\\
$^2$Department of Simulation Science, National Institute for Fusion Science, Oroshi--cho 322--6, Toki 509--5292, Japan.
}

\abst{
Energy density and energy flux was introduced along Takesue's method.
Particle energies were localized at particle positions using Dirac delta function.
The energy density was connected with the energy flux by continuity equation.
New method was proposed to deal with multi--body interaction.
Consequently, the energy current between particles could be calculated even if the multi--body interaction appeared.
Moreover, An application to the molecular dynamics simulation of hydrogen adsorption on a graphene with modified Brenner reactive empirical bond order potential was demonstrated.
}

\begin{document}

\maketitle

\section{Introduction}

In classical particle system, the energy current between particles was considered by Lepri, Livi and Politi with the approximation of low--$k$ limit \cite{Lepri}.
Takesue produced the energy current between particles from the energy density formed by the Dirac delta function and the continuity equation \cite{Takesue}.
This method is close to the work by Irving and Kirkwood \cite{IrvingKirkwood}.
These methods can treat only two--body interaction.
However, molecular dynamics (MD) which simulates atoms and molecules generally adopt multi--body interaction to represent molecular structures and chemical properties.
In the present paper, the Takesue's method is introduced and our new method enables to derive the energy current between particles from the multi--body interaction.

\section{Energy Field by Dirac Delta Function}

We derive the energy density and the energy flux according to Takesue's work\cite{Takesue}.
Hamiltonian in many--particle system is given by
\begin{align}
	H = \sum_i \frac{\nbk{\vc{p}_i(t)}^2}{2 m_i} + U(\vc{r}_1(t), \vc{r}_2(t), \cdots),
	\label{eq:2_1_1}
\end{align}
where $\vc{r}_i(t)$, $\vc{p}_i(t)$ and $m_i$ are the $i$--th particle position, momentum and mass, respectively.
Here, the $i$--th particle energy $e_i(t)$ is defined by 
\begin{align}
	e_i(t) = \frac{\nbk{\vc{p}_i(t)}^2}{2 m_i} + u_i(t),
	\label{eq:2_1_2}
\end{align}
where ${\nbk{\vc{p}_i(t)}^2}/{2m_i}$ and $u_i(t)$ are the $i$--th particle kinetic energy and interaction energy.
The particle interaction energies $u_i(t)$ satisfy
$ 
	U(\vc{r}_1(t), \vc{r}_2(t), \cdots) = \sum_i u_i(t).
$ 
Using the Dirac delta function, the particle energies $e_i(t)$ are localized at the particle positions $r_i(t)$.
Then the energy density $e(\vc{x},t)$ is defined by
\begin{align}
	e(\vc{x},t)=\sum_i e_i(t) \delta(\vc{x}-\vc{r}_i(t)),
	\label{eq:2_1_5}
\end{align}
where $\vc{x} = (x, y, z)$ is space coordinates.

Because total energy is a conservative quantity, the energy density constructs continuity equation 
\begin{align}
	\pdif{e(\vc{x},t)}{t} + \nabla \cdot \vc{j}(\vc{x},t) = 0,
	\label{eq:2_1_7}
\end{align}
with the vector field $\vc{j}(\vc{x},t)$ which is regarded as the energy flux.
The energy flux $\vc{j}(\vc{x},t)$ should consists of the following two parts.
As the $i$--th particle moves at velocity $\dot{\vc{r}}_i$, energy current $e_i \dot{\vc{r}}_i$ is generated at the $i$--th particle position $\vc{r}_i$.
In addition, interaction transports energy between particles by the interaction.
From these points of view, the energy flux $\vc{j}(\vc{x},t)$ is given by
\begin{align}
	\vc{j}(\vc{x},t)= \sum_i e_i(t) \dot{\vc{r}}_i(t) \delta(\vc{x}-\vc{r}_i(t))
	  + \sum_{i,k>i} j_{i \ra k}(t) \vc{\chi}_{ik}(\vc{x},t),
	\label{eq:2_1_10}
\end{align}
where $j_{i \ra k} (= - j_{k \ra i})$ is the magnitude of the energy current from the $i$--th particle to the $k$--th one and $\vc{\chi}_{ik}(\vc{x},t)$ is the vector field.
The first and second terms of the right--hand in Eq. (\ref{eq:2_1_10}) are the energy currents due to the movement of particles and due to interaction between particles, respectively.
To satisfy Eq. (\ref{eq:2_1_7}), the relation
\begin{align}
	\nabla \cdot \vc{\chi}_{ik}(\vc{x},t) = 
		\delta(\vc{x}-\vc{r}_i(t)) - \delta(\vc{x}-\vc{r}_k(t)),
	\label{eq:2_1_11}
\end{align}
is imposed on the vector field $\vc{\chi}_{ik}(\vc{x},t)$.
Thereby, we obtain
\begin{align}
	\dif{e_i(t)}{t} = - \sum_{k \neq i} j_{i \ra k}(t).
	\label{eq:2_1_15}
\end{align}
Thus, if the time derivative of the $i$--th particle energy ${\mathrm{d}e_i}/{\mathrm{d}t}$ consists of the elements of summation $\sum_{k \neq i}$, we can regard the elements as the energy current from $i$--th particle to the $k$--th one $j_{i \ra k}(t)$.
To derive $j_{i \ra k}(t)$ from Eq. (\ref{eq:2_1_15}), we must consider the $i$--th particle interaction energy $u_i(t)$.
Two--body interaction potential energy is simply allocated to the particle energies $u_i(t)$ into equal halves.
However, multi--body interaction potential energy cannot be divided into the the particle energies $u_i(t)$.

Determination the particle interaction energy $u_i(t)$ is unnecessary as long as the time derivative of the particle interaction energy ${\mathrm{d}u_i}/{\mathrm{d}t}$ is given.
If Hamiltonian system conserves total momentum, multi--body interaction potential has to be a function of relative position vectors
$ 
	 U(\vc{\bar{r}}_{12}(t), \cdots, \vc{\bar{r}}_{ik}(t), \cdots, \vc{\bar{r}}_{N-1,N}(t)),
$ 
where 
$ 
	\vc{\bar{r}}_{ik} \equiv \vc{r}_i - \vc{r}_k
$ 
is the relative position vector from the $k$--th particle to the $i$--th one.
We note that the multi--body interaction potential is not always the function only of the distance between particles $r_{ik} = \abs{\vc{\bar{r}}_{ik}}$.

According to the fact that the multi--body interaction potential is the function of the relative position vectors $\vc{\bar{r}}_{ik}$, the total differential of the total interaction potential is
\begin{align}
	 dU(\vc{\bar{r}}_{12}(t), \cdots, \vc{\bar{r}}_{ik}(t), \cdots, \vc{\bar{r}}_{N-1,N}(t))
		&= \sum_{i,k > i} \pdif{U}{\vc{\bar{r}}_{ik}} \cdot \mathrm{d} \vc{\bar{r}}_{ik}(t) \retn
		&= \frac{1}{2} \sum_{i,k \neq i} \pdif{U}{\vc{\bar{r}}_{ik}} \cdot \dot{\vc{\bar{r}}}_{ik}(t) \mathrm{d}t.
	\label{eq:2_24}
\end{align}
From this, the time derivative of the particle interaction energy is given by
\begin{align}
	 \dif{u_i}{t} = \frac{1}{2} \sum_{k \neq i} \pdif{U}{\vc{\bar{r}}_{ik}} \cdot \rbk{\dot{\vc{r}}_i(t) - \dot{\vc{r}}_k(t) }\mathrm{d}t.
	\label{eq:2_25}
\end{align}
Thereby, the time derivative of the $i$--th particle energy is written by the summation $\sum_{k \neq i}$ as follows;
\begin{align}
	\dif{e_i}{t} = \dif{}{t}\rbk{\frac{\nbk{\vc{p}_i}^2}{2m_i}} + \dif{u_i}{t}
	= - \frac{1}{2}\sum_{k \neq i} \rbk{\frac{\vc{p}_i}{m_i} + \frac{\vc{p}_k}{m_k} } \cdot \pdif{U}{\vc{\bar{r}}_{ik}}.
\label{eq:2_27}
\end{align}
As a result, the energy current between particles $j_{i \ra k}$ becomes
\begin{align}
	j_{i \ra k} = \frac{1}{2} \rbk{\frac{\vc{p}_i}{m_i} + \frac{\vc{p}_k}{m_k} } \cdot \pdif{U}{\vc{\bar{r}}_{ik}}.
\label{eq:2_28}
\end{align}
The partial derivative $\partial U / \partial {\vc{\bar{r}}_{ik}} (= - \partial {U} / \partial {\vc{\bar{r}}_{ki}})$ is the force which acts on the $k$--th particle due to the variation of the relative position vector $\vc{\bar{r}}_{ik}$ in the multi--body interaction.
This method is effective in arbitrary multi--body potentials.

\section{Application to Hydrogen Adsorption}

We demonstrate the application of the energy current between particles to the MD simulation which dealt with the adsorption of a hydrogen atom on a graphene.
The incident energy of the hydrogen atom was set to 3 eV.
Chemical interaction was represented by modified Brenner reactive empirical bond order potential \cite{Ito_ICNSP, Brenner}
\begin{align}
	U \equiv \sum_{i,j>i} \Bigg[V_{[ij]}^\mathrm{R}( r_{ij} ) - \bar{b}_{ij}(\{r\},\{\theta^\mathrm{B}\},\{\theta^\mathrm{DH}\}) V_{[ij]}^\mathrm{A}(r_{ij}) \Bigg],
	\label{eq:model_rebo}
\end{align}
where the functions $V_{[ij]}^{\mathrm{R}}$ and $V_{[ij]}^{\mathrm{A}}$ represent repulsion and attraction, respectively.
The function $\bar{b}_{ij}$ generates multi--body force where the bond angle $\theta^\mathrm{B}_{jik}$ and the dihedral angle $\theta^\mathrm{DH}_{kijl}$ are written by
$\cos\theta_{jik}^\mathrm{B} = {\vc{r}_{ji} \cdot \vc{r}_{ki}} / {r_{ji} r_{ki}}$
and 
$\cos\theta_{kijl}^\mathrm{DH} = \rbk{\vc{r}_{ik} \times \vc{r}_{ji}}
		\cdot \rbk{\vc{r}_{ji} \times \vc{r}_{lj}} / {r_{ik} r_{ji}^2 r_{lj}}$, respectively.

\begin{figure}
\begin{tabular}{cc}
		\resizebox{0.45\linewidth}{!}{\includegraphics{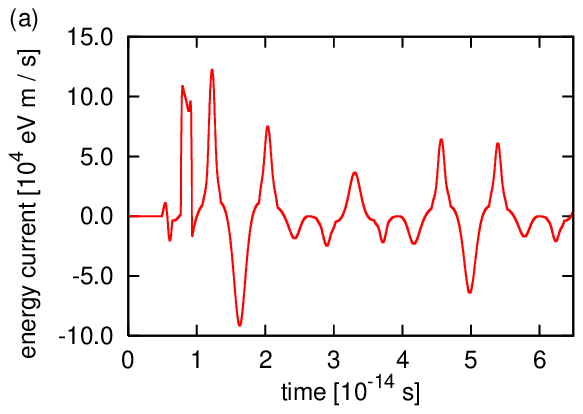}} &
		\resizebox{0.45\linewidth}{!}{\includegraphics{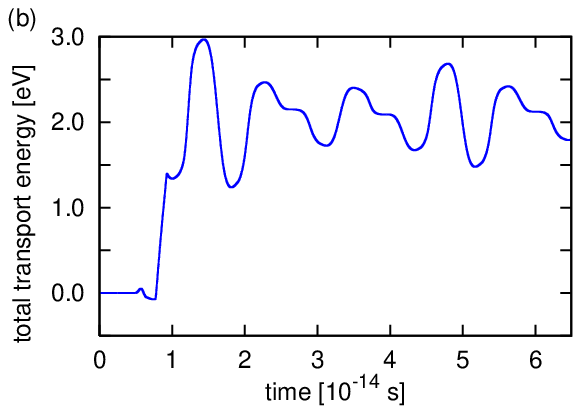}} \\
    \end{tabular}
\caption{(a) the energy current and (b) the total transport energy from the hydrogen atom to the graphene.}
\label{fig:XXX}
\end{figure}

The energy current from the hydrogen atom to the graphene
$j_\mathrm{H} \equiv \sum_{i \neq 1} j_{1 \rightarrow i}$
and total transport energy
$E_\mathrm{H} \equiv \int_0^t j_\mathrm{H} \mathrm{d}t$
were measured where the index of the hydrogen atom was 1.
Figure \ref{fig:XXX} shows $j_\mathrm{H}$ and $E_\mathrm{H}$ with time.
It was understood that the energy of about 2 eV was transferred from the hydrogen atom to the graphene as the hydrogen atom is adsorbed.

\section{Summary}

The energy density and the energy flux was introduced along the Takesue's method.
The particle energies $e_i(t)$ were localized at the particle positions $\vc{r}_i(t)$ using Dirac delta function as Eq. (\ref{eq:2_1_5}).
The energy density was connected with the energy flux by the continuity equation (\ref{eq:2_1_7}).
We proposed the new method to deal with the multi--body interaction.
The total differential of the multi--body interaction potential was expanded by not the particle positions $\vc{r}_i$ but the relative position vectors $\vc{\bar{r}}_{ik}$.
Consequently, the energy current between particles $j_{i \ra k}(t)$ was given by Eq. (\ref{eq:2_28}) in the multi--body interaction.
This method was applied to the MD simulation of the hydrogen adsorption on the graphene with modified Brenner REBO potential.

\section*{Acknowledgements}
The authors thank Dr. Shinji Takesue and Dr. Akira Ueda for helpful comments.
Numerical simulations were carried out by use of the Plasma Simulator at National Institute for Fusion Science.
The study was supported in part by a Grant--in--Aid for Exploratory Research (C), 2007, No. 17540384, from the Ministry of Education, Culture, Sports, Science and Technology, Japan, and in part by the National Institutes of Natural Sciences undertaking for Forming a Basis for Interdisciplinary and International Research through Cooperation Across Fields of Study, and Collaborative Research Programs (No. NIFS06KDAT012, NIFS06KTAT029, NIFS07USNN002, and NIFS07KEIN0091).

%

\end{document}